\documentclass[%
 twocolumn,
 superscriptaddress,
 amsmath,amssymb,
]{revtex4-2}

\usepackage[utf8]{inputenc}
\usepackage[OT1]{fontenc}
\usepackage[english]{babel}
\usepackage{soul}
\usepackage{siunitx}

\usepackage{graphicx}
  \graphicspath{{figures/}}
\usepackage{xcolor}

\usepackage{tabularx}

\usepackage{hyperref}

\begin{document}
\title{Acoustic Bound States in the Continuum in Coupled Helmholtz Resonators}

\author{Mariia Krasikova}
    \email{mariia.krasikova@metalab.ifmo.ru}
    \affiliation{School of Physics and Engineering, ITMO University, St. Petersburg 197101, Russia}
    \affiliation{Chair of Vibroacoustics of Vehicles and Machines, Technical University of Munich, Garching b. M\"unchen 85748, Germany}
\author{Felix Kronowetter}
    \thanks{These authors contributed equally.}
    \affiliation{Chair of Vibroacoustics of Vehicles and Machines, Technical University of Munich, Garching b. M\"unchen 85748, Germany}
\author{Sergey Krasikov}
    \thanks{These authors contributed equally.}
    \affiliation{School of Physics and Engineering, ITMO University, St. Petersburg 197101, Russia}
\author{Mikhail Kuzmin}
    \affiliation{School of Physics and Engineering, ITMO University, St. Petersburg 197101, Russia}
\author{Marcus Maeder}
    \affiliation{Chair of Vibroacoustics of Vehicles and Machines, Technical University of Munich, Garching b. M\"unchen 85748, Germany}
\author{Tao Yang}
    \affiliation{Chair of Vibroacoustics of Vehicles and Machines, Technical University of Munich, Garching b. M\"unchen 85748, Germany}
\author{Anton Melnikov}
    \affiliation{Chair of Vibroacoustics of Vehicles and Machines, Technical University of Munich, Garching b. M\"unchen 85748, Germany}
\author{Steffen Marburg}
    \affiliation{Chair of Vibroacoustics of Vehicles and Machines, Technical University of Munich, Garching b. M\"unchen 85748, Germany}
\author{Andrey Bogdanov}
    \email{a.bogdanov@metalab.ifmo.ru}
    \affiliation{School of Physics and Engineering, ITMO University, St. Petersburg 197101, Russia}
    \affiliation{Harbin Engineering University, Harbin 150001, Heilongjiang , Peoples R China}

\date{\today}
             
\begin{abstract}
Resonant states underlie a variety of metastructures that exhibit remarkable capabilities for effective control of acoustic waves at subwavelength scales. The development of metamaterials relies on the rigorous mode engineering providing the implementation of the desired properties. At the same time, the application of metamaterials is still limited as their building blocks are frequently characterized by complicated geometry and can't be tuned easily.
In this work, we consider a simple system of coupled Helmholtz resonators and study their properties associated with the tuning of coupling strength and symmetry breaking. We numerically and experimentally demonstrate the excitation of quasi-bound state in the continuum in the resonators placed in a free space and in a rectangular cavity. It is also shown that tuning the intrinsic losses via introducing porous inserts can lead to spectral splitting or merging of quasi-\textit{bound states in the continuum} and occurrence of \textit{exceptional points}. The obtained results will open new opportunities for the development of simple and easy-tunable metastructures based on Helmholtz resonances.
\end{abstract}

\maketitle

\section{Introduction}
Acoustic resonances play a pivotal role in a variety of practical applications covering all aspects of modern society, including art and entertainment~\cite{mcintyre1983oscillations,arns1995resonant,wolfe2009vocal}, medicine\mbox{~\cite{henderson1997essential,johns2002nonthermal,lange2008surface}}, industrial applications~\cite{leisure1997resonant,yuan2019recent}, and even cosmology~\cite{aerts2021probing}. The importance of resonances has become more prominent with the rise of metamaterials\mbox{~\cite{cummer2016controlling,ma2016acoustic,assouar2018acoustic}}, allowing effective control over acoustic fields at the subwavelength scale. Such remarkable properties as, for instance, negative mass density~\cite{lee2009acoustic,li2004doublenegative}, hyperlensing~\cite{li2009experimental}, or subwavelength perfect absorption~\cite{li2016acoustic} arise from the resonant states and interplay between them. Further progress in the field of metamaterials nowadays can be associated with resonant state engineering, involving such concepts as topological states~\cite{xue2022topological,zhang2023second}, non-locality~\cite{wang2023nonlocal}, and symmetry breaking~\cite{nassar2020nonreciprocity}. 

Of particular interest in this case are the so-called \textit{bound states in the continuum}~\cite{hsu2016bound,koshelev2023bound} (BIC), also known previously as trapped waves, which are characterized by infinitely large radiative quality factors and the associated decoupling from the far field, usually originating from the interference of several leaky modes. Firstly proposed in the field of quantum mechanics~\cite{vonneumann1993ueber}, BIC nowadays are actively studied in a variety of other systems, including electrical circuits~\cite{li2020bound}, photonics~\cite{koshelev2023bound,kang2023applications}, and acoustics~\cite{sadreev2021interference}. For instance, BIC can be utilized for sound confinement~\cite{huang2020extreme,deriy2022bound}, acoustic radiation enhancement~\cite{huang2024acoustic}, or improvement of noise-insulating systems~\cite{cao2021perfect,zhang2023topological}. Acoustic BIC were also studied in pipe and duct-cavity systems~\cite{hein2008acoustic,hein2012trapped}, as well as various configurations of single and coupled resonators~\cite{lyapina2015bound,huang2021sound,huang2022topological,huang2022general} and elastic systems~\cite{haq2021bound,cao2021elastic,lee2023elastic,an2024multibranch}. Since genuine BIC cannot be excited from the far field due to the decoupling, practical applications utilize the so-called quasi-BIC, which have a large but finite radiative quality factor. The excitation of quasi-BIC is associated with the symmetry breaking allowing weak coupling with the far field~\cite{koshelev2018asymmetric}.

\begin{figure}[htbp!]
    \centering
    \includegraphics[width=0.9\linewidth]{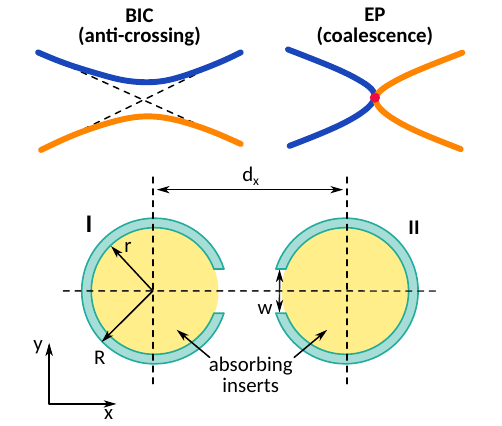}
    \caption{\textbf{Illustration of the work concept.} Bound states in the continuum (BIC) occur at the anti-crossings formed by real parts of symmetric and anti-symmetric modes in parametric space. Exceptional points (EP) represent a special case of coupling regime when the eigenmodes coalesce in the real and imaginary planes. In the considered system of two coupled Helmholtz resonators these regimes can be achieved by tuning the distance between the resonators and tuning their intrinsic losses. Both of the resonators are characterized by the inner radius $r$, the outer radius $R$ and the slit width $w$. In addition, the volume of the resonators is supplemented by absorbing inserts.}
    \label{fig:concept}
\end{figure}

Symmetry breaking plays a key role not only in the engineering of high-Q quasi-BICs but also in bianisotropic~\cite{asadchy2018bianisotropic,craig2019experimental,melnikov2019acoustic,li2020bianisotropic,melnikov2022microacoustic} and non-Hermitian systems~\cite{ashida2020nonhermitian,wang2023nonlocal,huang2024acoustic}. In particular, \textit{exceptional point} (EP), which denotes a condition where two or more eigenmodes and their complex eigenfrequencies coalesce, is achieved via breaking of parity (P-symmetry) and time-reversal symmetry (T-symmetry)~\cite{heiss2012physics,ozdemir2019parity,igoshin2024exceptional}. EPs find today a variety of applications in photonics and acoustics~\cite{miri2019exceptional,huang2024acoustic}. Practical implementations of EP in acoustic systems have enabled asymmetric field propagation~\cite{shen2018synthetic}, mode stabilization~\cite{bourquard2019stabilization}, and enhanced absorption~\cite{achilleos2017nonhermitian}.

Therefore, it can be inferred that such singularities as BIC and EP are pivotal for the engineering of acoustics metamaterials and devices on their base with unique functionality and performance. However, their practical implementations are still limited by the typically nontrivial geometry of unit cells and the narrow operational frequency range. Producing metastructures with easily tunable designs and acoustic properties is also a difficult task, sometimes requiring the use of non-trivial algorithms of numerical optimization. Therefore, the creation of resonant meta-atoms with straightforward designs that facilitate the easy adjustment of resonant and singular states, including BIC and EP, could greatly enhance the adoption of metamaterials in real-world applications.

In this study, we suggest and explore a straightforward design of meta-atom with easily tuned acoustic properties. The suggested meta-atom 
represents a pair of two-dimensional coupled C-shaped Helmholtz resonators. Such a system can be tuned via mechanical displacement
of the resonators to each other as well as via the adjustment of their intrinsic losses using absorbing inserts made of conventional porous materials. We demonstrate experimentally that such a meta-atom supports quasi-BIC that could be merged into an EP via the tuning of intrinsic losses.

\section{Results}
\subsection{System description}
The considered system consists of coupled 2D Helmholtz resonators with circular cross-sections, as shown in Fig.~\ref{fig:concept}. The coupling can be tuned via the change of distance between the resonators and their mutual orientation, while the intrinsic losses can be controlled by absorbing inserts. Resonators of similar shapes were considered, for instance, in Refs.~\cite{melnikov2019acoustic,melnikov2020acoustic,kronowetter2023sound,krasikova2023metahouse,krasikova2024broadband}. Normal modes of such a system are formed by the superposition of the modes of solitary resonators. In-phase excitation of the resonators corresponds to a symmetric mode, which in optics is usually called a bright one since it can be directly excited and observed from the far field. On the other hand, the resonators can be excited out-of-phase, forming an anti-symmetric mode. 

When an antisymmetric mode is localized entirely inside the resonators, its radiative quality factor is infinite. Such states are genuine BIC, whose coupling with the far field is prohibited by symmetry reasons. Therefore, the excitation of BIC implies a symmetry break, allowing the transformation of BIC into a quasi-BIC, whose radiative quality factor is still large but finite. Regarding eigenmodes, BIC occur in the strong mode coupling regime when the dispersion curves (real parts of eigenfrequencies) repel. For the weak coupling, the situation is reversed and the real parts of modes are crossed. In between these regimes, achieving an exceptional point (EP) in which both real and imaginary parts of the modes coalesce and the eigenmodes become linearly dependent. In the considered system, the coupling between the modes can be easily tuned via simple displacement of the resonators, affecting the radiative losses. At the same time, the intrinsic losses and coupling between the resonators can be tuned via the introduction of porous inserts into their volume. Hence, it might be expected that a simple variation of several parameters is enough to achieve the considered coupling regimes.

\subsection{Quasi-BIC}
\subsubsection{Numerical Simulations}
\begin{figure*}[htbp!]
    \centering
    \includegraphics[width=0.9\linewidth]{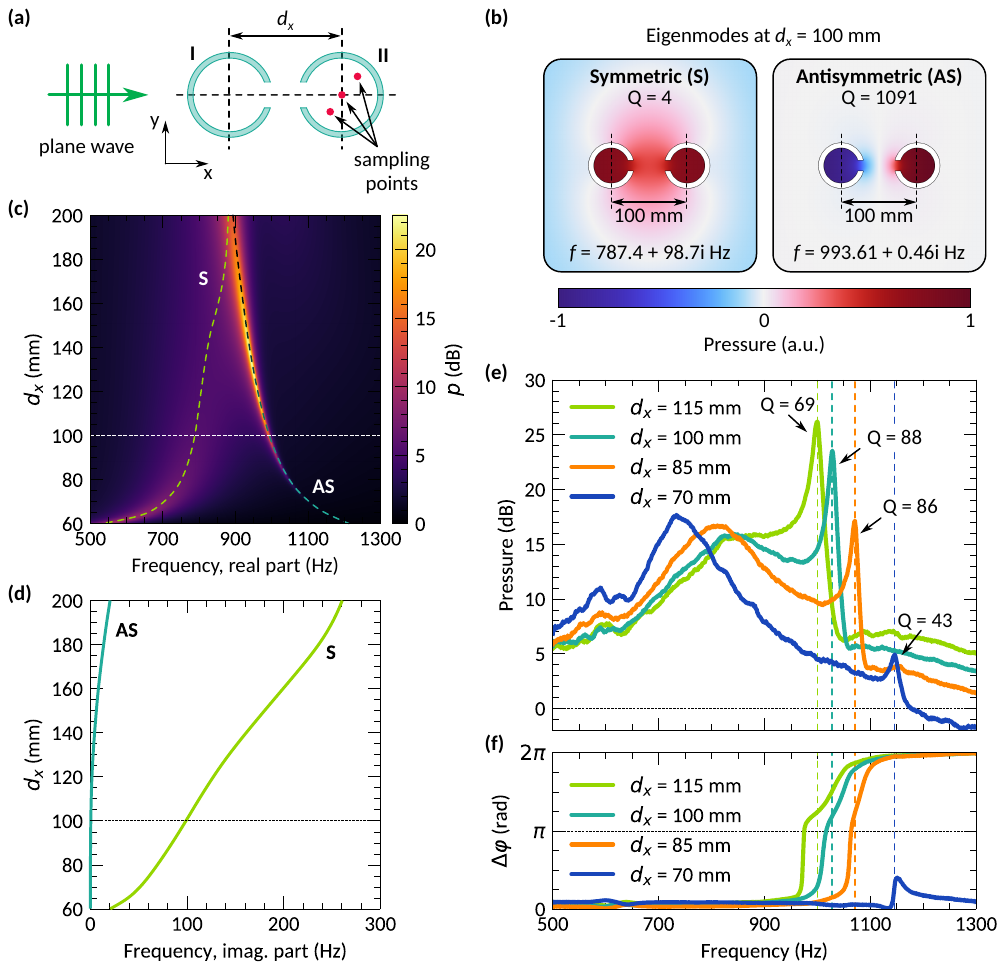}
    \caption{\textbf{Quasi-BIC in the resonators placed in a free space.} (a) Schematic illustration of the considered system. (b) Eigenmodes field distributions at $d_x = \SI{100}{\milli\meter}$ and their quality factor $Q$. (c) Pressure spectra inside one of the resonators as a function of the distance $d_x$ between the resonators. The dashed lines indicate the corresponding eigenmodes with symmetric ($S$) and anti-symmetric ($AS$) field distributions. (d) Imaginary part of the eigenfrequencies as a function of $d_x$. (e) Measured pressure spectra inside one of the resonators. The values of $Q$ indicate the corresponding quality factors of the resonances. (f) Measured spectra of phase difference between the fields inside the resonators.}
    \label{fig:open_spectra_dx}
\end{figure*}

The considerations start with the resonators without absorbing inserts. For this case, the geometric parameters of the resonators were fixed to have the outer radius of the resonators $R = \SI{30}{\milli\meter}$, the inner radius $r = \SI{23}{\milli\meter}$, and the slit width $w = \SI{14}{\milli\meter}$, which correspond to the scaled geometry considered in Refs.~\cite{krasikova2023metahouse,krasikova2024broadband}. At first, the resonators placed in a free space are considered [see Fig.~\ref{fig:open_spectra_dx}(a)]. Such a system is characterized by two eigenmodes with symmetric and anti-symmetric field distributions [Fig.~\ref{fig:open_spectra_dx}(b)], which correspond to the resonances of pressure spectra inside one of the resonators [Fig.~\ref{fig:open_spectra_dx}(c)]. When the distance between the resonators becomes large enough, they almost do not interact and their eigenfrequency become almost identical corresponding to the frequency of a single Helmholtz resonator. On the contrary, the spectral width of the mode with the anti-symmetric distribution vanishes for small $d_x$ when resonators are close to each other, which is associated with the vanishing of the imaginary parts of eigenmodes [Fig.~\ref{fig:open_spectra_dx}(d)]. Therefore, the real parts of the modes merge with the increase of $d_x$, demonstrating the independent behavior of two non interacting resonators, while the imaginary parts diverge and vice versa. This is a manifestation of a symmetry-protected quasi-BIC, which becomes a genuine BIC when the imaginary part of the mode is equal to zero. 

\begin{figure*}[htbp!]
    \centering
    \includegraphics[width=0.9\linewidth]{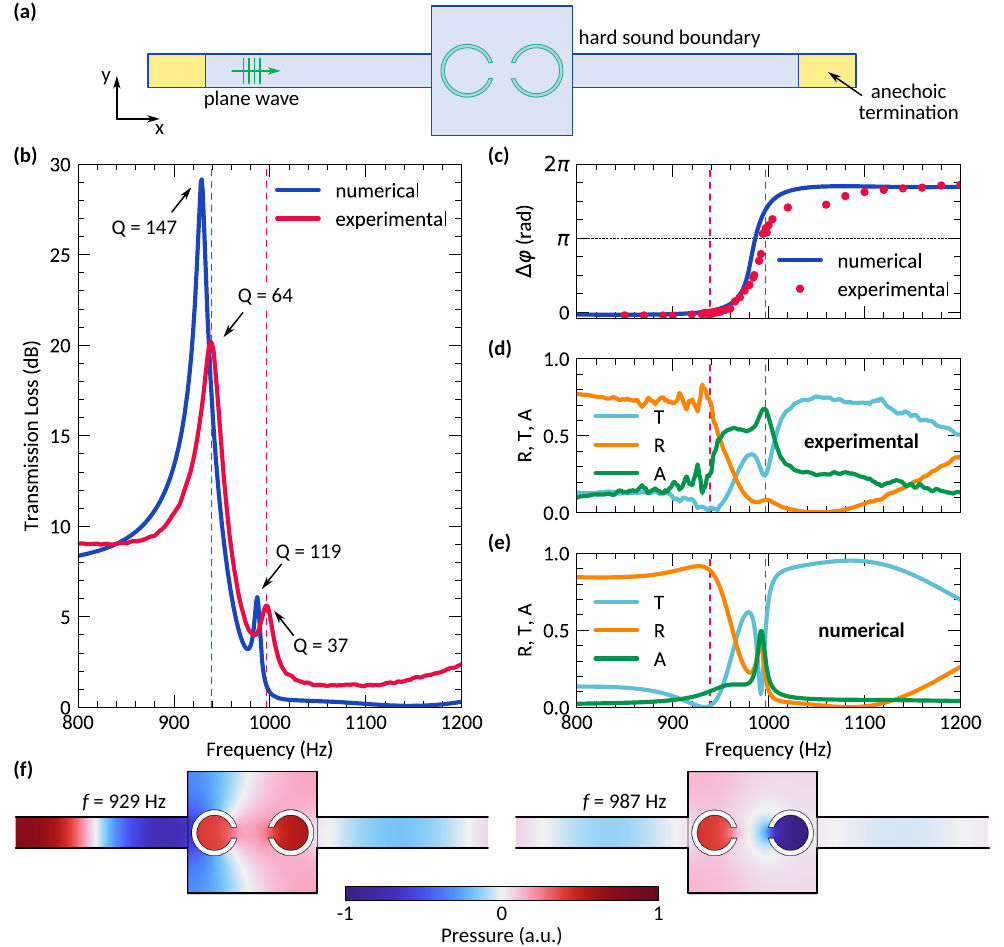}
    \caption{\textbf{Transmission tube measurements of quasi-BIC.} (a) Schematic illustration of the system consisting of the box with the resonators embedded into the transmission tube. (b) Transmission loss spectra. (c) Experimentally measured and numerically calculated spectra of phase difference between the fields inside the resonators. (d) Experimentally obtained and (e) numerically calculated reflection, transmission and absorption coefficients. (f) Field distributions corresponding to the transmission loss resonances occurring near \SI{929}{\hertz} and \SI{987}{\hertz}.}
    \label{fig:tube_dx100}
\end{figure*}

\subsubsection{Anechoic Chamber Measurements}
The numerical calculations are verified by the measurements of pressure inside the resonators placed in the 2D anechoic chamber (see Methods). 
As it is shown in [Fig.~\ref{fig:open_spectra_dx}(e)], the quasi-BIC manifests as a peak in the pressure spectra, occurring within the range $900-\SI{1200}{\hertz}$, depending on the distance between the resonators. For instance, when $d_x = \SI{100}{\milli\meter}$ the quasi-BIC is excited at approximately $\SI{1025}{\hertz}$, and its estimated Q-factor is $88$. The comparison of the frequencies and the Q-factors is provided in Table~\ref{tab:open_spectra_Q}, while the explicit comparison of the numerical and experimental spectra is shown in Supplementary Information. In accordance with the numerical calculations, the quasi-BIC shifts towards the higher frequencies with a decrease in the distance between the resonators since the coupling between the resonators becomes stronger. At the same time, decrease of the distance results in the weakening of the coupling with the far field, so the corresponding resonance can't be excited properly meaning that its amplitude and Q-factor are small, which is observed for both numerical and experimental results.

\begin{table}[htbp!]
    \centering
    \caption{Quality factor of quasi-BIC resonances of the system consisting of two coupled resonators placed in a free space.}
    \label{tab:open_spectra_Q}
    \begin{tabular}{c|cc|cc}
        & \multicolumn{2}{c|}{Frequency, Hz} & \multicolumn{2}{c}{Q-factor} \\
        \cline{2-5}
        $d_x$, mm & Num. & Exp. & Num. & Exp. \\
        \hline
        70 & 1101 & 1146 & 95 & 43 \\
        85 & 1029 & 1071 & 137 & 86 \\
        100 & 991 & 1028 & 131 & 88 \\
        115 & 967 & 1000 & 107 & 69
    \end{tabular}
\end{table}

By definition, the anti-symmetric mode is characterized by the $\pi$ phase difference between the fields inside the resonators, which can be seen from the spectra in [Fig.~\ref{fig:open_spectra_dx}(f)], demonstrating the conventional smoothstep-like shapes. 
Note that for the case of $d_x = \SI{70}{\milli\meter}$, the phase difference does not reach the value of $\pi$, which might be explained by the weak coupling of the antisymmetric mode with the far field. 

\begin{figure*}[htbp!]
    \centering
    \includegraphics[width=0.9\linewidth]{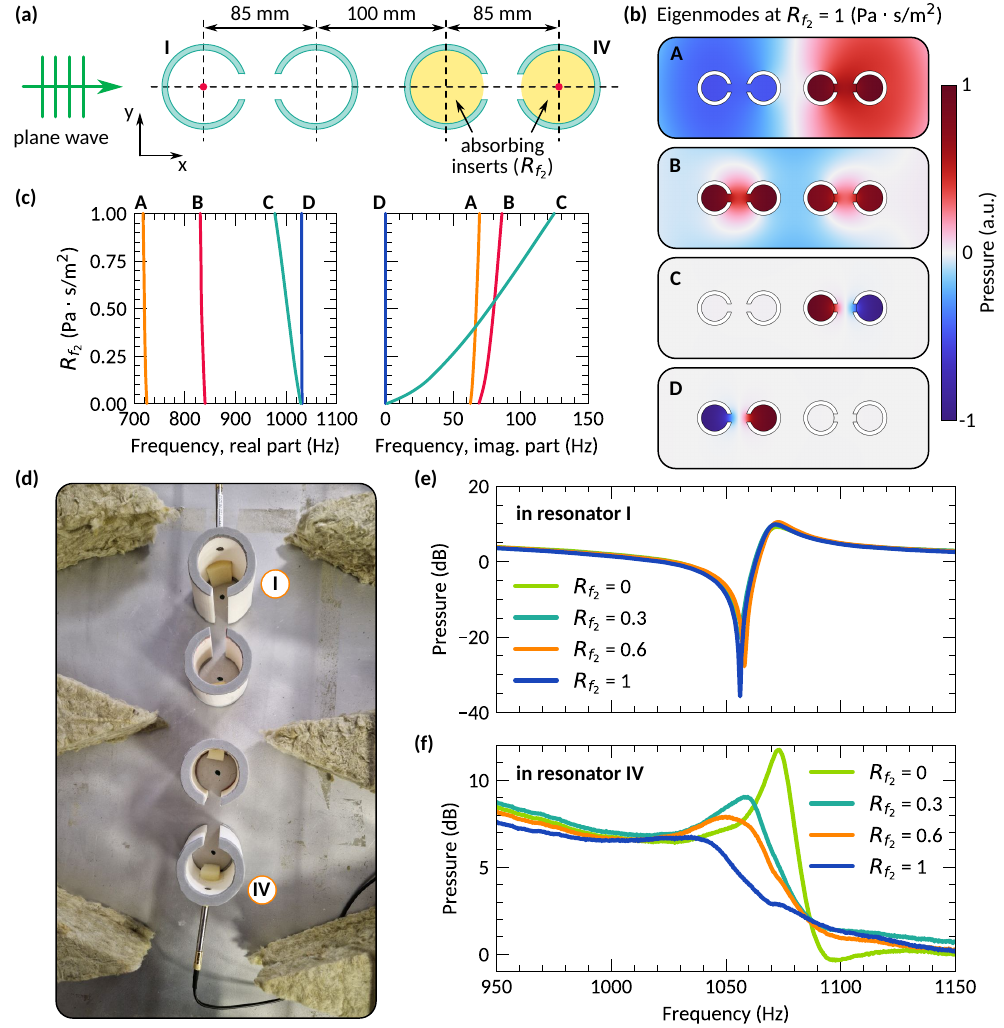}
    \caption{\textbf{Splitting of quasi-BIC} (a) Schematic illustration of the considered system consisting of two pairs of coupled resonators placed in a free space. One of the pairs is supplemented by porous inserts with the flow resistivity $R_{f_2}$. (b) Eigenmodes field distributions for the case of $R_{f_2} = \SI{1}{\pascal\cdot\second\per\meter^2}$ and (c) real and imaginary parts of eigenmodes for different values of $R_{f_2}$. (d) Photo of the samples. Experimentally obtained pressure spectra measured inside (e) the resonator I and (f) the resonator IV.}
    \label{fig:open_eigenmodes_BIC_splitting}
\end{figure*}

In addition, it should be noted that the resonances of the measured spectra are shifted from the numerically calculated ones by approximately \SI{30}{\hertz}--\SI{50}{\hertz}, which arises from manufacturing errors as sub-millimeter changes in the resonators sizes affect the position of the resonances (see Supplementary Information). However, the experimental results are in good agreement with those from numerical simulations. It also should be mentioned that apart from the geometric parameters the position of quasi-BIC is affected by material parameters of the media in which the resonators are placed. Such a property might be promising for the development of devices for sensing of gaseous and liquid analytes (see Supplementary Information for an example).

\subsubsection{Transmission Tube Measurements}
While the quasi-BIC mode is weakly coupled to the far field, it is also almost decoupled from the other modes of the system. Hence, it can be expected that the quasi-BIC can still be excited even if the environment of the resonators changes. For instance, the resonators can be placed inside a rectangular cavity [see Fig.~\ref{fig:tube_dx100}(a)], which actually by itself supports quasi-BIC~\cite{kronowetter2023realistic}. 
When the anti-symmetric modes of the cavity and the resonators do not interact, the transmission loss spectra of such a system should also demonstrate a high-Q peak near the frequency corresponding to the quasi-BIC occurring for the case of the resonators in a free space. Indeed, Fig.~\ref{fig:tube_dx100}(b) shows the presence of the two pronounced resonances near \SI{940}{\hertz} and \SI{1000}{\hertz}, obtained via the impedance tube measurements (see Methods). Again, the phase difference between the fields inside the resonators represents the smooth step-like curve [see Fig.~\ref{fig:tube_dx100}(c)], demonstrating that the phase difference is nearly $\Delta\varphi = \pi$ at the quasi-BIC frequency. 

The peak of the transmission loss spectrum occurring near \SI{940}{\hertz} can be associated with the symmetric excitation of Helmholtz resonators [see Fig.~\ref{fig:tube_dx100}(f)], leading to the decrease of the transmission coefficient [Fig.~\ref{fig:tube_dx100}(d)].
At the same time, the second peak in the transmission spectra, occurring near \SI{1000}{\hertz}, corresponds to the quasi-BIC characterized by the anti-symmetric field distribution, as previously [see Fig.~\ref{fig:tube_dx100}(f)].  The resonance in the transmission loss, in this case, is related to the increase of the absorption coefficient [see Figs.~\ref{fig:tube_dx100}(d) and ~\ref{fig:tube_dx100}(e)]. However, it should be noted that numerical and experimental results in this case do not perfectly agree with each other. While in both cases there is and absorption peak at the frequency of quasi-BIC, for the experimental curve the absorption is higher, especially at the frequency of the symmetric resonance. Such a difference is associated with energy leakage into the mechanical vibrations of the box walls, which are not taken into account in the numerical calculations. The presence of the box wall vibrations are verified using the laser Doppler vibrometry technique (Polytec PSV-500-3D), but elimination or analysis of this vibrations lie beyond the scope of the present work. 

\begin{figure*}[htbp!]
    \centering
    \includegraphics[width=0.9\linewidth]{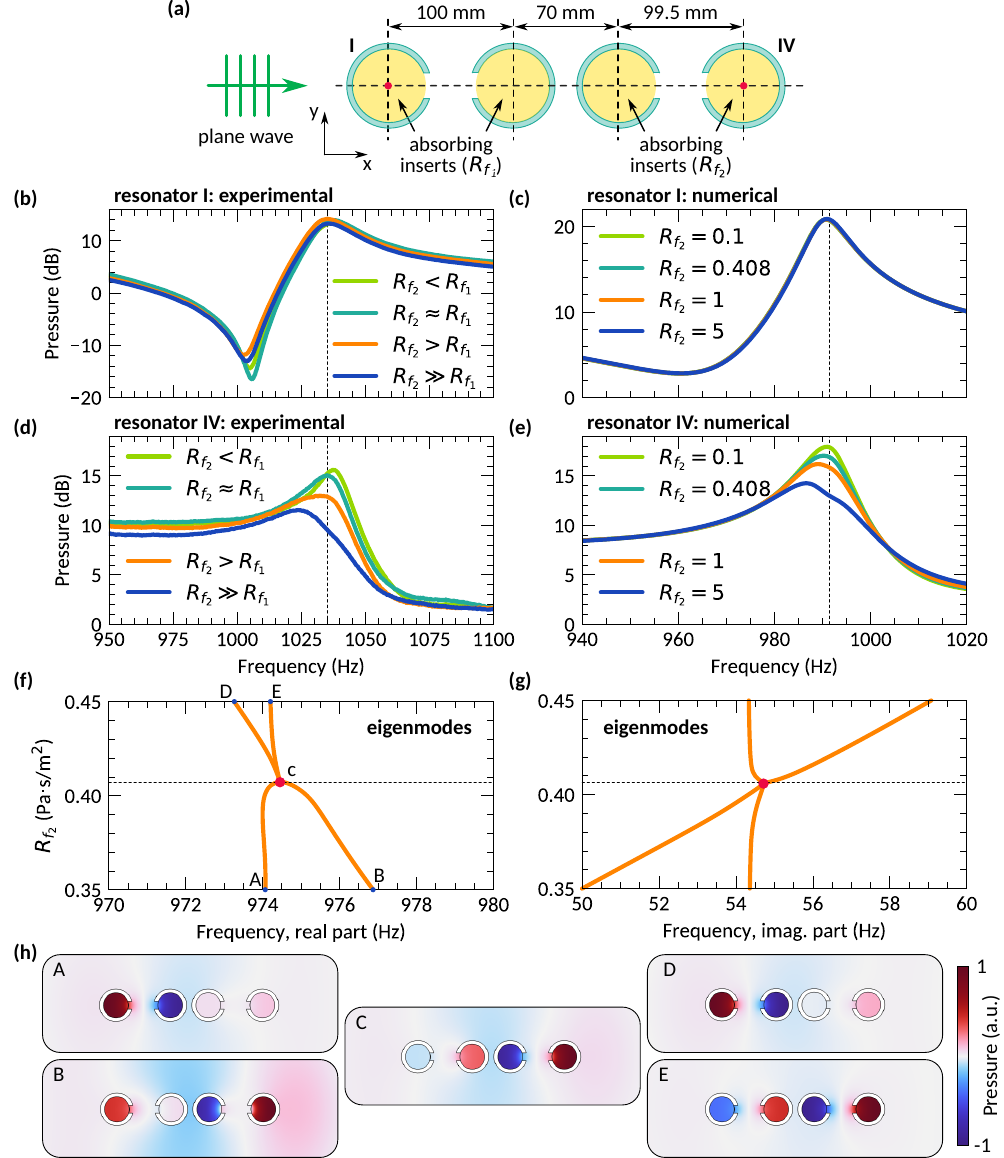}
    \caption{\textbf{EP in the system of four resonators.} (a) Schematic illustration of the considered system consisting of two pairs of coupled resonators, both supplemented by porous inserts with the flow resistivity $R_{f_1}$ and $R_{f_2}$, respectively. Pressure spectra experimentally measured inside (b) the resonator I and (d) the resonator IV, and the corresponding spectra numerically calculated inside (c) the resonator I and (e) the resonator IV (sampling points are shown by the red dots). All numerical results are obtained for $R_{f_1} = \SI{0.4}{\pascal\cdot\second\per\meter^2}$. (f) The real and (g) imaginary parts of eigenmodes for the different values of $R_{f_2}$, and (h) the corresponding field distributions.}
    \label{fig:open_eigenmodes_EP}
\end{figure*}

\subsection{Tuning of Intrinsic Losses}

Apart from the distance between the resonators, the resonances can be controlled by introducing porous materials into the volume of the resonators. Such inserts may significantly decrease the quality factor of a standalone Helmholtz resonator (see Supplementary Information), practically leading to the destruction of the resonance. At the same time, introduction of intrinsic losses may result in the inversion symmetry break and the associated asymmetric absorption (see Supplementary Information). Small intrinsic losses can be used for the fine-tuning of quasi-BIC.
For instance, the system supplemented by an additional pair of the resonators can be considered, such that this additional pair is supplemented by the absorbing inserts made of a porous material with the flow resistivity $R_{f_2}$, while another pair has no such inserts [see Fig.~\ref{fig:open_eigenmodes_BIC_splitting}(a)]. 
Since there are four resonators in such a system, it is characterized by four eigenmodes [see Figs.~\ref{fig:open_eigenmodes_BIC_splitting}(b)], two of which are characterized by the symmetric field distributions within the pairs and two -- by the antisymmetric [see Figs.~\ref{fig:open_eigenmodes_BIC_splitting}(c)]. 
When the flow resistivity $R_{f_2}$ is zero, the antisymmetric eigenmodes are nearly degenerate, meaning that the have equal real parts of the frequencies, but slightly different values for the imaginary parts. Both the real and imaginary parts of the symmetric modes are almost not affected by the change of $R_{f_2}$. At the same time, the increase of $R_{f_2}$ results in the splitting of quasi-BIC [see Figs.~\ref{fig:open_eigenmodes_BIC_splitting}(b) and~\ref{fig:open_eigenmodes_BIC_splitting}(c)], such that the frequency of one antisymmetric eigenmode remains the same, while another antisymmetric eigenmode shifts towards the smaller real frequencies, while its quality factor quickly decreases. 
Similar behaviour can be also observed experimentally. In this case the spectral position of the quasi-BIC measured in the first pair remains nearly unchanged [see Fig.~\ref{fig:open_eigenmodes_BIC_splitting}(e)], while the quasi-BIC measured in the second pair shifts towards the lower frequencies with the increase of the absorbing insert size [see Fig.~\ref{fig:open_eigenmodes_BIC_splitting}(d)]. These results indicate the spectral positions of BIC in arrays of Helmholtz resonators can be fine-tuned via simple adjustment of the intrinsic losses controlled by the size of the absorbing inserts.

The important remark is that in the numerical calculations, the insert occupies the whole volume of the resonator, and the corresponding porous material is characterized only by the flow resistivity, which is varied. For the experimental verification, it is assumed that a small piece of porous material with a large flow resistivity is roughly equivalent to a large piece of a material with a low flow resistivity. In such an approximation, a porous insert occupying the whole volume of the resonator can be represented by an insert occupying only a part of the volume [see Fig.~\ref{fig:open_eigenmodes_BIC_splitting}(d)], and it is much easier to control the size of an insert contrary to its material parameters.

To push the concept further, it should be mentioned that two quasi-BIC can be merged into an EP~\cite{qin2022exceptional}.
Strictly speaking, an EP is characterized by coalescence of eigenmodes rather than their degeneracy~\cite{heiss2012physics}, which implies intersection of both real and imaginary parts of the eigenmodes in parametric space. This condition can be achieved via fine-tuning of intrinsic losses of the resonators and their coupling~\cite{ding2016emergence}. For that, again, absorbing inserts can be used, such that in the first pair, the inserts are characterized by the flow resistivity $R_{f_1}$ and in the second pair -- by $R_{f_2}$ [see Fig.~\ref{fig:open_eigenmodes_EP}(a)]. When losses of the first pair are fixed, the corresponding quasi-BIC excited in the first pair remains unchanged [see Figs.~\ref{fig:open_eigenmodes_EP}(b) and~\ref{fig:open_eigenmodes_EP}(c)]. The change of $R_{f_2}$ however results in the deformation of quasi-BIC formed in the second pair, such that its frequency shifts towards the lower frequencies [see Figs.~\ref{fig:open_eigenmodes_EP}(d) and~\ref{fig:open_eigenmodes_EP}(e)]. At some point, resonances occurring in the first and the second pairs become spectrally overlapped, which might indicate the presence of an EP. Indeed, fine-tuning of intrinsic losses and the coupling between the resonators results in the spectral merging of two eigenmodes in both real and imaginary spaces [see Figs.~\ref{fig:open_eigenmodes_EP}(f)--~\ref{fig:open_eigenmodes_EP}(h)], similarly to other systems of four coupled resonators~\cite{peng2024regulation}.

It should be mentioned however, that the spectral positions of the resonances do not fully correspond to the eigenfrequencies. This happens due to the fact that the system is characterized by four eigenmodes having rather large imaginary parts, meaning that the corresponding spectral widths of the resonances are also large and they may overlap. In particular, not only quasi-BIC but also a symmetric mode contribute to the observed resonances. This is also the reason why the resonance with $R_{f_2} \gg R_{f_1}$ looks to be slightly deformed, as increase of $R_{f_2}$ results in the interaction between the symmetric and antisymmetric modes. A comprehensive analysis of eigenmodes and EPs however lies beyond the scope of the work the aim of which is to develop a simply tunable metaatom allowing to control quasi-BIC.

\begin{figure*}[htbp!]
    \centering
    \includegraphics[width=0.9\linewidth]{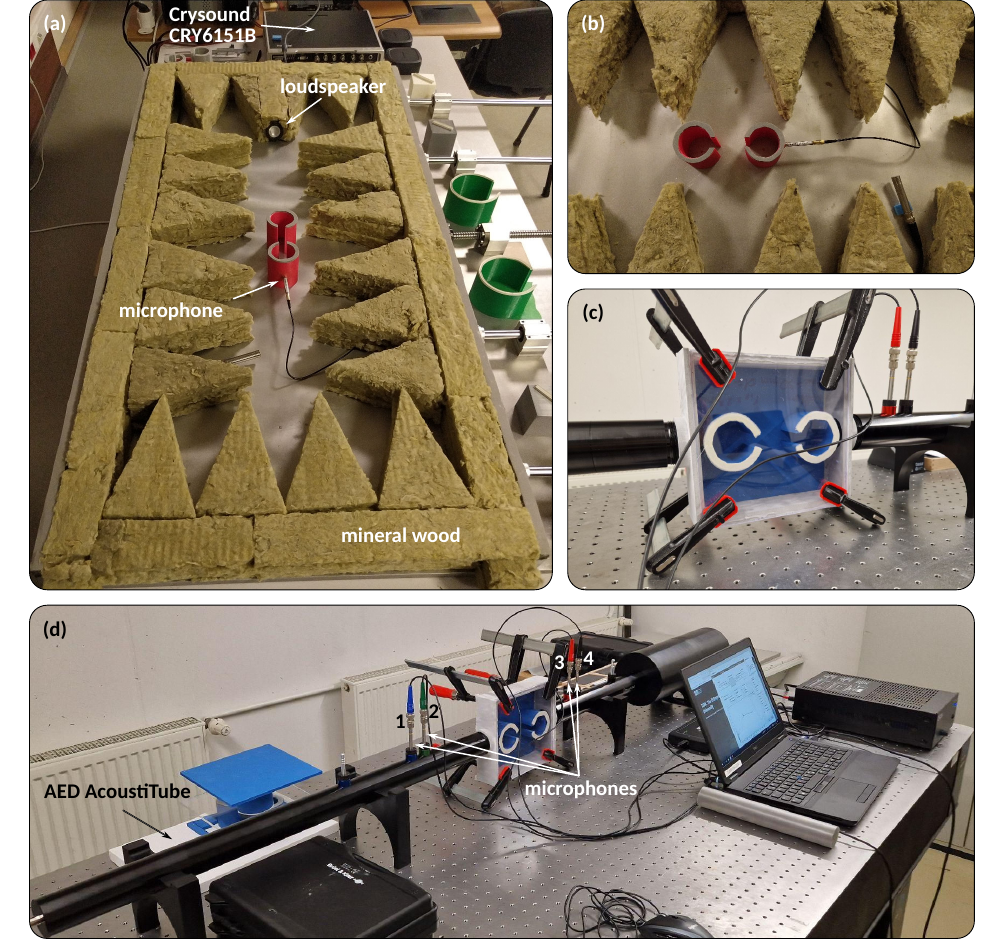}
    \caption{\textbf{Experimental Setups and Samples.} (a) Photo of the 2D anechoic chamber imitating a free space. Close up photo of (b) the samples used in the 2D anechoic chamber and (c) the samples and the box used for the transmission tube measurements. (d) Photo of the transmission tube setup.}
    \label{fig:methods}
\end{figure*}

\section{Conclusion}
In this work, coupling regimes of 2D Helmholtz resonators were investigated. It was shown that quasi-BIC and EP can be easily achieved by tuning the geometric parameters and intrinsic losses of the resonators. Experimental demonstrations reveal the resonant enhancement of absorption associated with quasi-BIC.
The obtained results aim to extend the possibilities of physics-based design of metamaterials and pave the route towards novel devices for the efficient control of acoustic fields at the subwavelength scales.

\section{Methods}
\subsection{Experimental measurements}
For the resonators placed in a free space, the measurements are performed in the 2D anechoic chamber, representing a waveguide with walls covered by absorbing mineral wood [see Figs.~\ref{fig:methods}(a) and~\ref{fig:methods}(b)]. In particular, the walls and the bottom of the waveguide are made of \SI{15}{\milli\meter} thick aluminum plates, and the lid is made of \SI{6}{\milli\meter} plexiglass. Since the impedance contrast between air and most of the solid materials is huge ~\cite{selfridge1985approximate}, the lid and the bottom can be considered as sound hard walls. The height of the waveguide is \SI{60}{\milli\meter}, while the width and the length are \SI{650}{\milli\meter} and \SI{1350}{\milli\meter}, respectively. The thickness of the mineral wood inserts is \SI{300}{\milli\meter} (from a wall to the tip of a pyramid). Generation and recording of the sound signal is done using the analysis system Crysound CRY6151B and 1/4" pressure field microphone Crysound CRY342. In order to mitigate the influence of random noises, the signal is recorded several times and then averaged in the Fourier domain. 

The transmission loss measurements are conducted in the AED AcoustiTube measurement system, the impedance tube. In this case, the resonators are placed in a box with a rectangular cross-section [see Fig.~\ref{fig:methods}(c) and ~\ref{fig:methods}(d)]. The width of the box is \SI{170}{\milli\meter}, the height is \SI{160}{\milli\meter}, and the thickness is \SI{40}{\milli\meter}. The scattering coefficients are extracted from the measured T-matrix coefficients using the conventional expressions~\cite{jimenez2021acoustic} as:
\begin{gather}
    t^{-} = \frac{2 e^{ikL}}{T_{11} + T_{12}/Z_L + T_{21} Z_0 + T_22 Z_0/Z_L},
    \\
    t^{+} = \frac{Z_0}{Z_L} \frac{2 e^{ikL}(T_{11}T_{22} - T_{12} T_{21})}{T_{11} + T_{12}/Z_L + T_{21} Z_0 + T_22 Z_0/Z_L},
    \\
    r^{-} = \frac{T_{11} + T_{12}/Z_L - T_{21} Z_0 - T_22 Z_0/Z_L}{T_{11} + T_{12}/Z_L + T_{21} Z_0 + T_22 Z_0/Z_L},
    \\
    r^{+} = \frac{-T_{11} + T_{12}/Z_L - T_{21} Z_0 + T_22 Z_0/Z_L}{T_{11} + T_{12}/Z_L + T_{21} Z_0 + T_22 Z_0/Z_L},    
\end{gather}
where $Z_0$, $Z_L$ are acoustic impedances of the media at the sides of the structure, such that $L$ is the thickness of the structure, and the superscripts indicate the positive ($+$) and ($-$) negative direction of the incident wave propagation. The amplitude coefficients then can be defined as $T^{+,-} = |t^{+,-}|^2$ and $R^{+,-} = |r^{+,-}|^2$. Correspondingly, the absorption coefficients can be defined as $A^{+,-} = 1 - |r^{+,-}|^2 - |t^{+,-}|^2$. When the system is symmetric, $R = R^{+} = R^{-}$, $T = T^{+} = T^{-}$, $A = A^{+} = A^{-}$.
Note that all measured spectra are slightly smoothed using the Savitzky-Golay filter of order $3$ and with a window width of $11$.

All resonators are fabricated by 3D printing using the well-known fused deposition modeling (FDM) technology, for which the polylactic acid (PLA) is used as a filament.  In the case of 2D anechoic chamber measurements, the resonators are manufactured on the Creality K1 3D printer (Shenzhen Creality 3D Technology Co., Ltd., Shenzhen, China).  For the impedance tube measurements, the Bambu Lab X1-Carbon 3D printer (Shenzhen Tuozhu Technology Co., Ltd., Shenzhen, China) is used to produce the resonators. Typical \SI{0.4}{\milli\meter} nozzles and standard slicing settings are used for both 3D printers to achieve the sample resolution of approximately \SI{0.2}{\milli\meter}. To avoid gaps between the resonators and the lids, the top of the resonators are supplemented by thin porous inserts.

The quality factor of the quasi-BIC is estimated from the spectra curves as the ratio of the center frequency to the $\SI{-3}{\deci\bel}$ bandwidth:
\begin{equation}
    Q = \dfrac{f_0}{\Delta f_{\SI{-3}{\deci\bel}}}.
\end{equation}
The central frequency $f_0$ is determined programmatically as the frequency at which the maxima of the peak is located. The value of amplitude at $f_0$ is considered as $y_0$ and the bandwidth is determined as the difference of frequencies at which the resonance curve intersect the $y_0 - \SI{3}{\deci\bel}$ line (the corresponding sketch is provided in the Supplementary Information). Note that for the case of $d_x = \SI{70}{\milli\meter}$ the peak is not pronounced and there is only one intersection of the resonance curve and the $y_0 - \SI{3}{\deci\bel}$ line. Hence, this particular resonance curve is fitted using Lorentzian function and bandwidth is calculated for this fitting curve. 

\subsection{Numerical calculations}
All numerical calculations are performed in COMSOL Multiphysics using the ``Pressure Acoustics'' physics. The incident plane wave is introduced as the background pressure field with the amplitude $p_0 = \SI{1}{\pascal}$. 
The pressure inside the resonators is defined as a root mean square of pressure calculated at several sampling points located inside the resonators.
In order to account for the thermoviscous dissipation in the boundary layer of the resonators' walls, the ``Thermoviscous Boundary Layer Impedance'' condition is applied to the corresponding boundaries. Porous inserts are included via the ``Poroacoustics'' feature utilizing the Delany-Bazley-Miki model. 

For the case of transmission tube, the 3D model is considered, in which the transmission coefficient is calculated as the amplitude of the total pressure over the corresponding background pressure, i.e.
\begin{equation}
    T = \frac{P_t}{P_b},
\end{equation}
such that the pressure amplitudes are averaged over a small volume in the transmission zone of the structure:
\begin{equation}
    P_{t,b} = \frac{1}{V} \int\limits_{V} |p_{t,b}|^2 dV.
\end{equation}
The corresponding transmission loss coefficient is then obtained as 
\begin{equation}
    TL = 10 \mathrm{log}_{10}(1/T).
\end{equation}

\begin{acknowledgments}
M.Kr., S.K. and A.B. acknowledges the BASIS foundation. 
\end{acknowledgments}

\section{Author contributions}
M.Kr., M.K., F.K., M.M., and T.Y. performed the experimental measurements, M.Kr., S.K. and F.K. provided numerical calculations. S.K. and M.Kr. suggested the idea and developed the concept of the work. A.M., M.M., S.M. and A.B. supervised the project. All authors contributed to writing and
editing of the manuscript.

\section{Competing interests}
The authors declare no competing interests.

\bibliographystyle{unsrt}
\bibliography{references.bib}

\end{document}


\title{Supplementary Information for \\ Acoustic Bound States in the Continuum in Coupled Helmholtz Resonators}

\author{Mariia Krasikova}
    \email{mariia.krasikova@metalab.ifmo.ru}
    \affiliation{School of Physics and Engineering, ITMO University, St. Petersburg 197101, Russia}
    \affiliation{Chair of Vibroacoustics of Vehicles and Machines, Technical University of Munich, Garching b. M\"unchen 85748, Germany}
\author{Felix Kronowetter}
    \thanks{These authors contributed equally.}
    \affiliation{Chair of Vibroacoustics of Vehicles and Machines, Technical University of Munich, Garching b. M\"unchen 85748, Germany}
\author{Sergey Krasikov}
    \thanks{These authors contributed equally.}
    \affiliation{School of Physics and Engineering, ITMO University, St. Petersburg 197101, Russia}
\author{Mikhail Kuzmin}
    \affiliation{School of Physics and Engineering, ITMO University, St. Petersburg 197101, Russia}
\author{Marcus Maeder}
    \affiliation{Chair of Vibroacoustics of Vehicles and Machines, Technical University of Munich, Garching b. M\"unchen 85748, Germany}
\author{Tao Yang}
    \affiliation{Chair of Vibroacoustics of Vehicles and Machines, Technical University of Munich, Garching b. M\"unchen 85748, Germany}
\author{Anton Melnikov}
    \affiliation{Chair of Vibroacoustics of Vehicles and Machines, Technical University of Munich, Garching b. M\"unchen 85748, Germany}
\author{Steffen Marburg}
    \affiliation{Chair of Vibroacoustics of Vehicles and Machines, Technical University of Munich, Garching b. M\"unchen 85748, Germany}
\author{Andrey Bogdanov}
    \affiliation{School of Physics and Engineering, ITMO University, St. Petersburg 197101, Russia}
    \affiliation{Harbin Engineering University, Harbin 150001, Heilongjiang , Peoples R China}

\date{\today}

\maketitle

\tableofcontents

\clearpage
\section{Comparison of the Experimental and Numerical Results}
\begin{figure*}[htb!]
    \centering
    \includegraphics[width=0.9\linewidth]{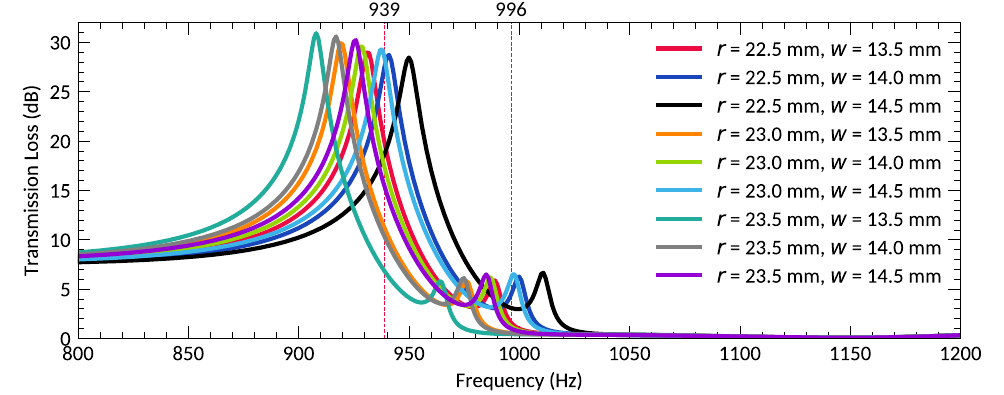}
    \caption{\textbf{The influence of the geometric parameters of the resonators on numerical calculations}. Transmission loss spectra for the case of two coupled resonators place in the rectangular box. The distance between the resonators is $d_x = \SI{100}{\milli\metre}$ and the outer radius is $R = \SI{30}{\milli\metre}$, while the inner radius $r$ and the slit width $w$ are varied. The vertical dashed lines indicate the position of peaks in the experimentally measured spectra.}
    \label{fig:tube_TL_spectra_dx100_geom_error}
\end{figure*}

The experimental results presented in the main text are in a good agreement with the numerical calculations. However, the resonant peaks in the pressure and transmission loss spectra are noticeably shifted with respect to each other. Such a difference can be explained by the limited accuracy of the samples manufacturing. For instance, even slight inaccuracies of order \SI{0.5}{\milli\metre} may lead to the significant shift of the transmission loss resonances, as shown in Fig.~\ref{fig:tube_TL_spectra_dx100_geom_error}. The corresponding measured spectrum lies within the range of the numerical curves, and hence it can be concluded that the experimental results reproduce the numerical ones.

\begin{figure*}[htbp!]
    \centering
    \includegraphics[width=0.9\linewidth]{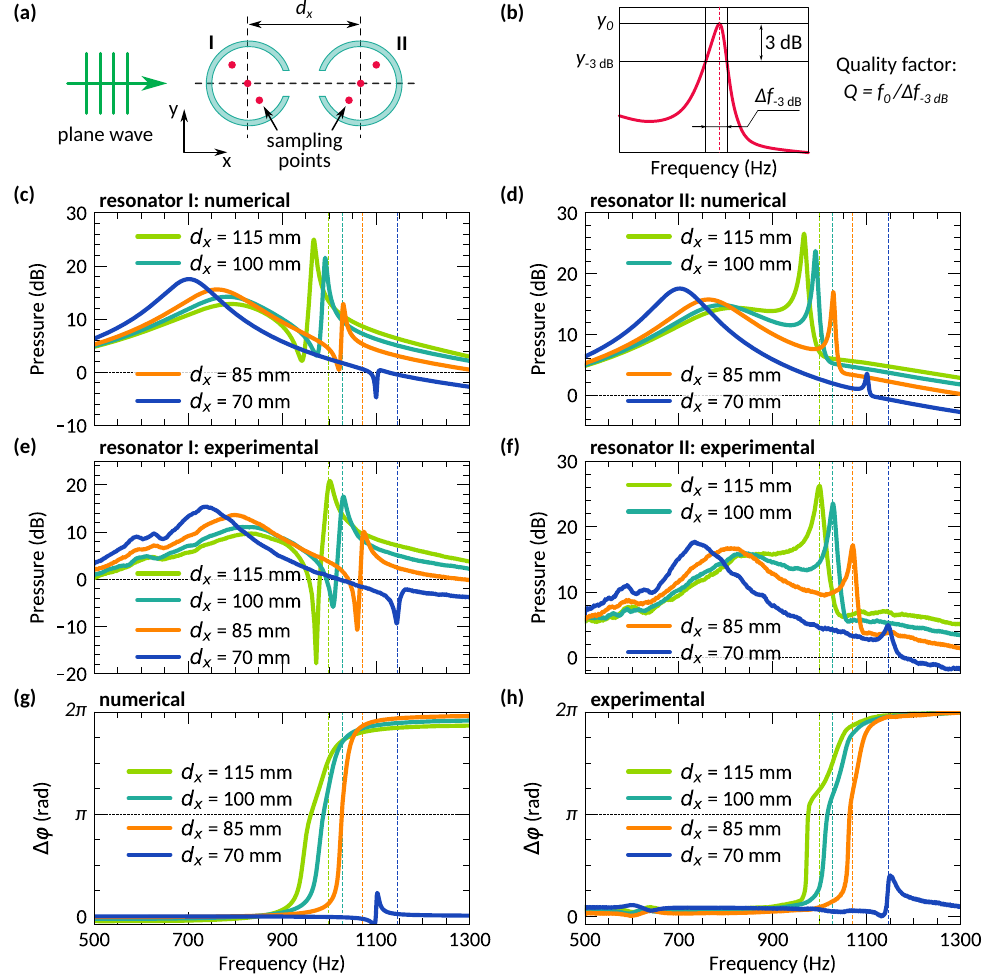}
    \caption{\textbf{Pressure spectra in the open system of coupled resonators.} (a) Schematic illustration of the system consisting of two resonators excited by a plane wave. (b) Schematics of the Q-factor estimation. The pressure spectra (c) calculated and (e) measured inside the resonator I and the corresponding spectra (d) calculated and (f) measured inside the resonator II. (g) Numerical and (h) experimental phase difference of the fields inside the resonators. Vertical lines correspond to the experimentally measured frequencies of quasi-BIC.}
    \label{fig:open_spectra_dx_left_right}
\end{figure*}

It is also should be mentioned that when a pair of the resonators is considered [see Figs.~\ref{fig:open_spectra_dx_left_right}(a)] the pressure spectra inside them are slightly different. In particular, the pressure spectra measured and calculated inside one of the resonators are characterized by Fano-shaped resonances corresponding to the antisymmetric mode [see Figs.~\ref{fig:open_spectra_dx_left_right}(b) and~\ref{fig:open_spectra_dx_left_right}(d)], while for the second resonator the corresponding peaks have Lorentzian-like shape [see Figs.~\ref{fig:open_spectra_dx_left_right}(c) and~\ref{fig:open_spectra_dx_left_right}(e)]. 
The corresponding quality factors of the resonances, provided in the main text, are estimated as the ratio of the resonant frequency over the $\SI{-3}{\deci\bel}$ bandwidth, which are determined in accordance to the sketch shown in Fig.~\ref{fig:open_spectra_dx_left_right}(b). The estimated numerical and experimental phase differences of the fields inside the resonators also follow the same trend. While for the distances $85$, $100$ and $\SI{115}{\milli\meter}$ the excitation of quasi-BIC corresponds to the phase difference $\pi$, for the case of $d_x = \SI{70}{\milli\meter}$ the resonance is not properly excited and hence the phase difference does not reach $\pi$.

\begin{figure}[htbp!]
    \centering
    \includegraphics[width=0.9\linewidth]{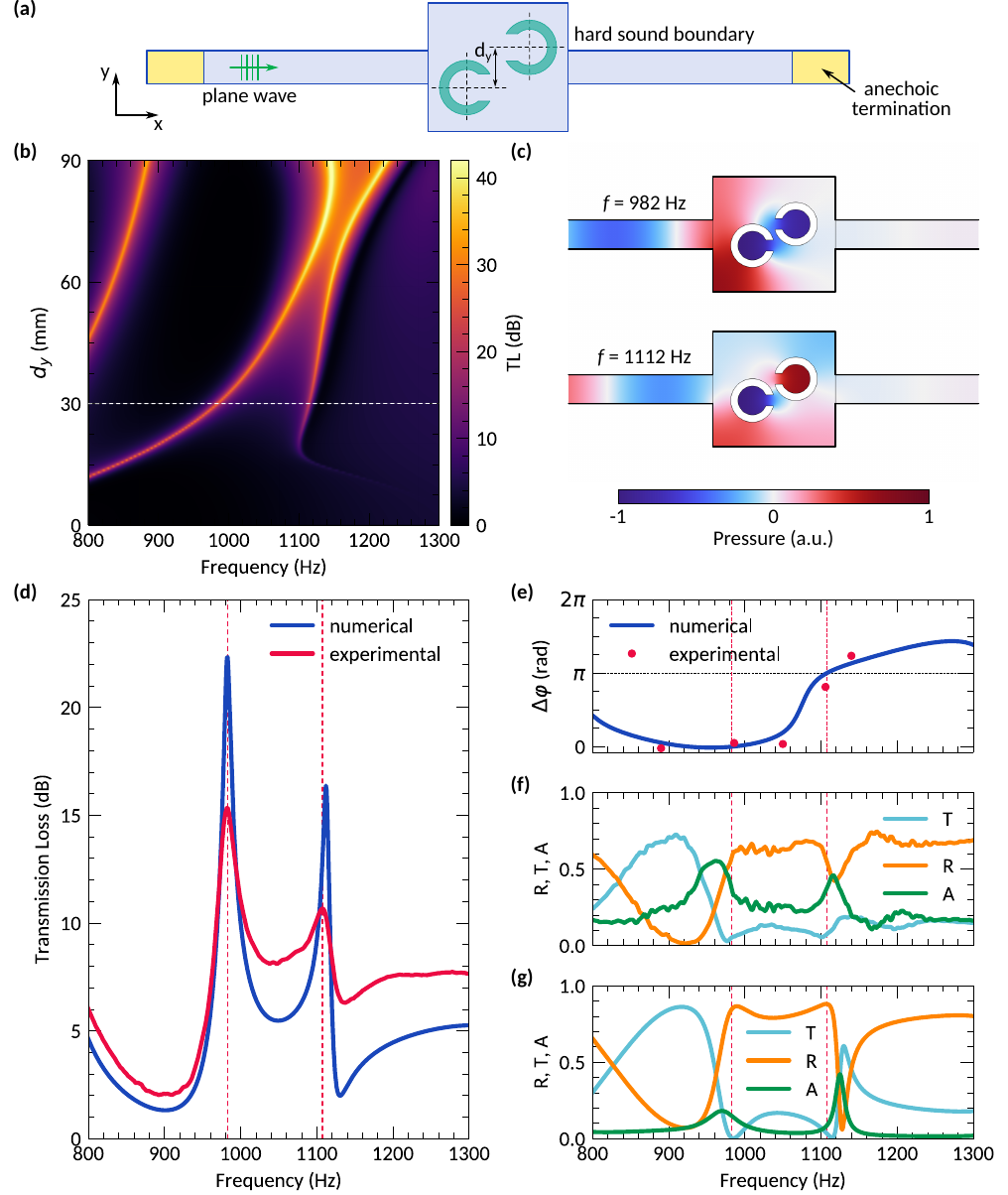}
    \caption{\textbf{Breaking of geometric symmetry.} (a) Schematic illustration of the system consisting of the box with the resonators embedded into the transmission tube. (b) Transmission loss spectra as a function of the distance between the resonators $d_y$. 
    (c) Field distributions corresponding to the transmission loss resonances occurring near \SI{982}{\hertz} and \SI{1112}{\hertz} for the case of $d_y = \SI{30}{mm}$. (d) Comparison of the experimentally measured and numerically calculated transmission loss spectra for the case of $d_y = \SI{30}{mm}$. (e) Experimentally obtained and (f) numerically calculated reflection, transmission and absorption coefficients.}
    \label{fig:tube_phase_spectra_dy30}
\end{figure}

It also should be noted that the symmetric and antisymmetric resonances lie rather close to each other. In order to make the spectral separation large, the distance between the resonators should be adjusted. In addition, it is reasonable to consider how the displacement of the resonators along the direction perpendicular to the wave propagation [see Fig.~\ref{fig:tube_phase_spectra_dy30}(a)] will affect the transmission and absorption spectra. To avoid overlaps between the modes associated with the Helmholtz resonators and the modes of the box in which they are placed, in this case the geometry of the resonators is slightly changed, such that the inner radius $r = \SI{20}{\milli\meter}$, the slit width is $w = \SI{15}{\milli\meter}$ and the distance between the resonators along the $x$-axis is $d_x = \SI{65}{\milli\meter}$. As shown in Fig.~\ref{fig:tube_phase_spectra_dy30}(b), the increase of $d_y$ is associated with the decrease of the spectral distance between the resonances. Note, that even for rather large values of $d_y$, for instance $d_y = \SI{30}{\milli\meter}$, the antisymmetric mode still exist [see Fig.~\ref{fig:tube_phase_spectra_dy30}(b)] and the associated transmission loss peak is well-distinguished in the spectra [see Fig.~\ref{fig:tube_phase_spectra_dy30}(c)]. As previously, the excitation of the antisymmetric mode can be traced via determination of the phase difference of the fields inside the resonators, such that for the antisymmetric mode it should be near $\pi$. Indeed, this condition is satisfied for both the numerical and experimental results [see Fig.~\ref{fig:tube_phase_spectra_dy30}(d)], and similarly to the case considered in the main text, the excitation of the symmetric and antisymmetric modes is associated with the increase of absorption coefficient [see Fig.~\ref{fig:tube_phase_spectra_dy30}(e) and ~\ref{fig:tube_phase_spectra_dy30}(f)].

\clearpage
\section{Tuning of Intrinsic Losses}
\begin{figure*}[htbp!]
    \centering
    \includegraphics[width=0.9\linewidth]{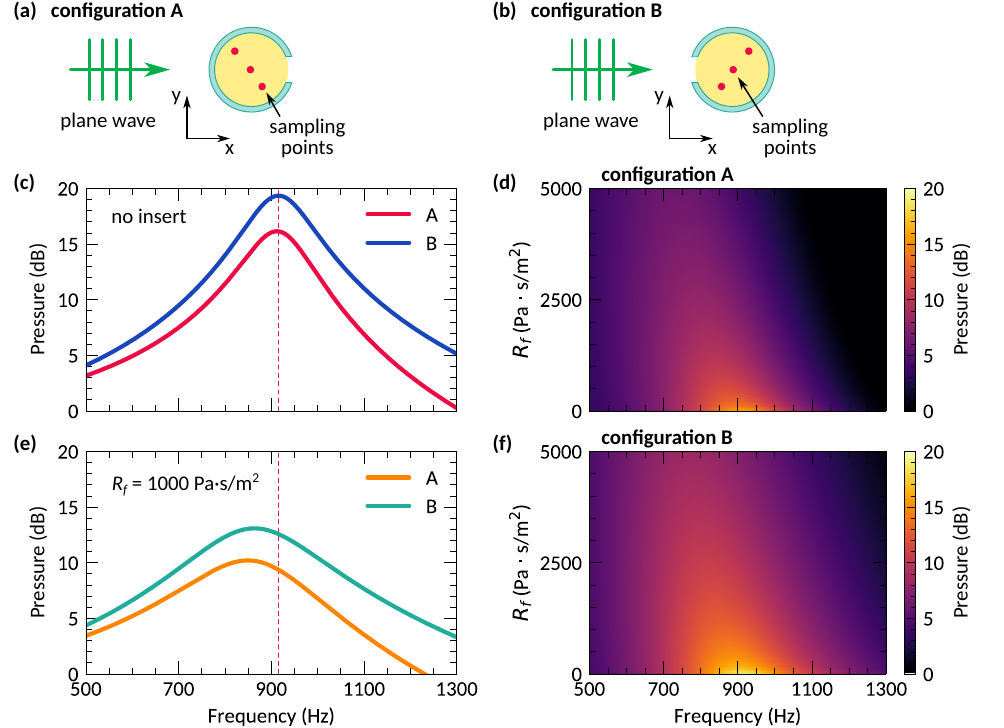}
    \caption{\textbf{Standalone resonator.} Schematic geometry of the open system with a standalone resonator oriented (a) ``away'' from (configuration A) and (b) ``towards'' (configuration B) the incident field. The porous insert with the flow resistivity $R_f$ are indicated by the yelow color.  Pressure spectra calculated inside the resonators (c) without porous inserts and (e) with the inserts characterized by the flow resistivity . The vertical red lines indicate the pressure peak corresponding to the configuration A without the insert. Pressure spectra calculated inside the resonators as functions of the flow resistivity of the porous inserts for the configurations (d) A and (f) B.}
    \label{fig:open_spectra_single}
\end{figure*}

While the main text is dedicated to the systems of coupled resonators, it is still reasonable to analyse a standalone resonator. In particular, two configurations are considered, such that the resonator is oriented ``away'' from and ``towards'' the incident field [see Fig.~\ref{fig:open_spectra_single}(a) and~\ref{fig:open_spectra_single}(b)]. In both cases the resonance occurs near $900$~Hz, though for the configuration A the pressure is slightly than for the B [see Fig.~\ref{fig:open_spectra_single}(c)]. When the resonator is supplemented by a porous insert, the spectral width of the resonance increases with the increase of the flow resistivity, while the pressure decreases [see Fig.~\ref{fig:open_spectra_single}(d) --~\ref{fig:open_spectra_single}(f)]. 

Large values of the flow resistivity practically result in the disappearance of the resonance as the intrinsic losses become too large and the pressure spectra flattens. Hence, even for the case of a standalone resonator it is reasonable to consider small values of flow resistivity, which can be achieved via considerations of the partial inserts, occupying only a small part of the resonator's volume, as discussed in the main text.

\begin{figure*}[htbp!]
    \centering
    \includegraphics[width=0.9\linewidth]{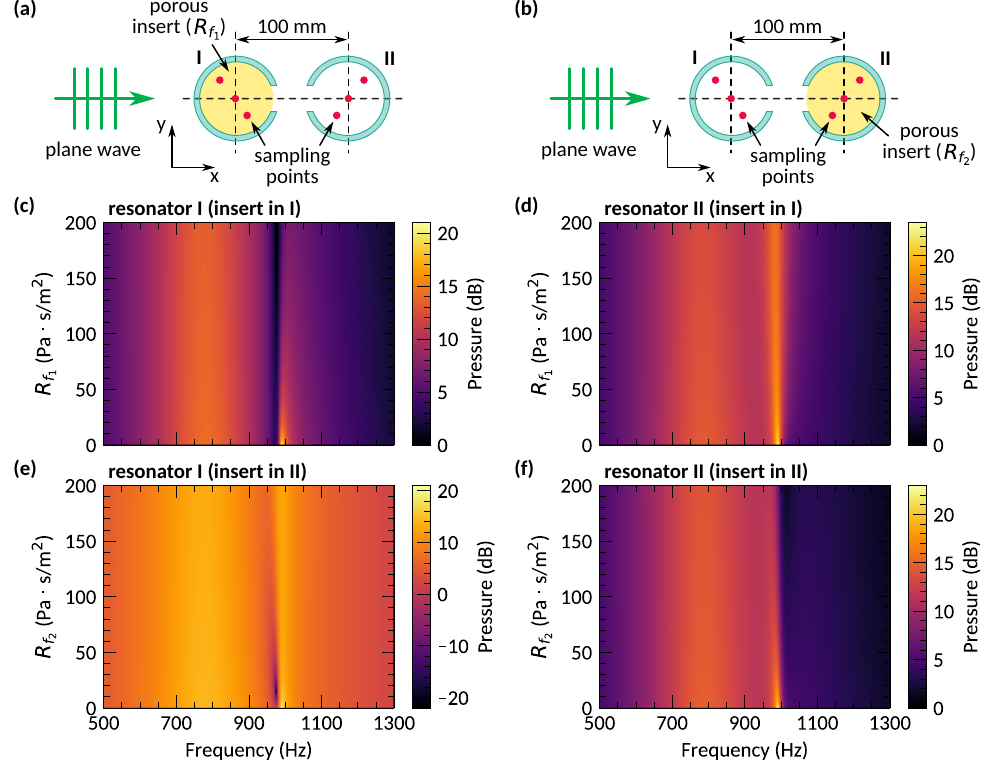}
    \caption{\textbf{Tuning of losses in one of the resonators within a pair.} Schematic illustration of the considered open system consisting of two resonators, such that the porous insert is placed inside (a) the resonator I and (b) the resonator II. Pressure spectra for the case when the insert is placed in the resonator I, calculated inside (c) the resonator I and (d) the resonator II. Pressure spectra for the case when the insert is placed in the resonator II, calculated inside (e) the resonator I and (f) the resonator II. }
    \label{fig:open_spectra_dx100_porous_left_right}
\end{figure*}

When a second resonator is introduced into the system, the coupling between the resonatorsresults in the splitting of the resonance. As discussed in the main text, one part of the split resonance corresponds to the symmetric mode and shifts towards the lower frequencies with the increase of the coupling strength. The second resonance corresponds to the antisymmetric mode having larger frequency and larger quality factor than the symmetric one. When a porous insert is introduced into one of the resonators within the pair [see Figs.~\ref{fig:open_spectra_dx100_porous_left_right}(a) and~\ref{fig:open_spectra_dx100_porous_left_right}(b)], the amplitude of the quasi-BIC in the resonator with the insert quickly decreases with the increase of the flow resistivity [see Figs.~\ref{fig:open_spectra_dx100_porous_left_right}(c) and~\ref{fig:open_spectra_dx100_porous_left_right}(f)]. At the same time, the change of the corresponding amplitude in the resonator without the insert does not change so drastically [see Figs.~\ref{fig:open_spectra_dx100_porous_left_right}(d) and~\ref{fig:open_spectra_dx100_porous_left_right}(e)]. In other words, the increase of intrinsic losses in one of the resonators results in the asymmetric excitation of the resonators, as it should be expected.

\begin{figure*}[htbp!]
    \centering
    \includegraphics[width=0.9\linewidth]{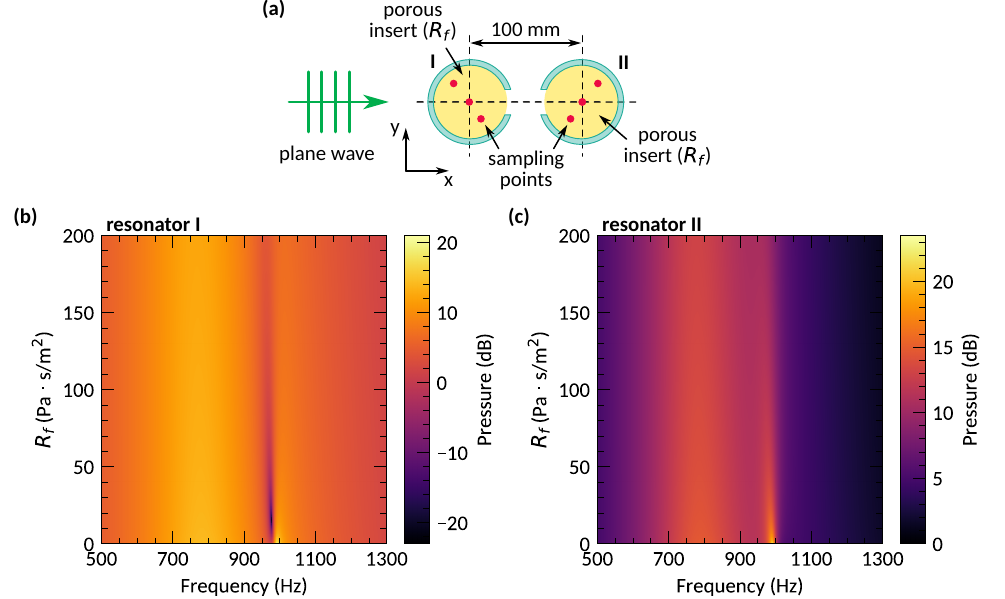}
    \caption{\textbf{Tuning of losses in both resonators within a pair.} (a) Schematic illustration of the considered open system consisting of the two resonators supplemented by porous inserts with the flow resistivity $R_f$. Pressure spectra calculated inside (b) the resonator I and (c) the resonator~II.}
    \label{fig:open_spectra_dx100_porous_both}
\end{figure*}

In the case when the inserts are introduced into the both resonators [see Fig.~\ref{fig:open_spectra_dx100_porous_both}(a)], the resonance amplitude also decreases with the increase of the flow resistivity, such that at $R_f = 200$~Pa$\cdot$s/m$^2$ the antisymmetric mode becomes indistinguishable. Therefore, tuning of the quasi-BIC via intrinsic losses should be done with materials characterized by the effective flow resistivity below $10$~Pa$\cdot$s/m$^2$ or even $1$~Pa$\cdot$s/m$^2$. 

\clearpage
\section{Inversion Symmetry Break}
\begin{figure}[htbp!]
    \centering
    \includegraphics[width=0.9\linewidth]{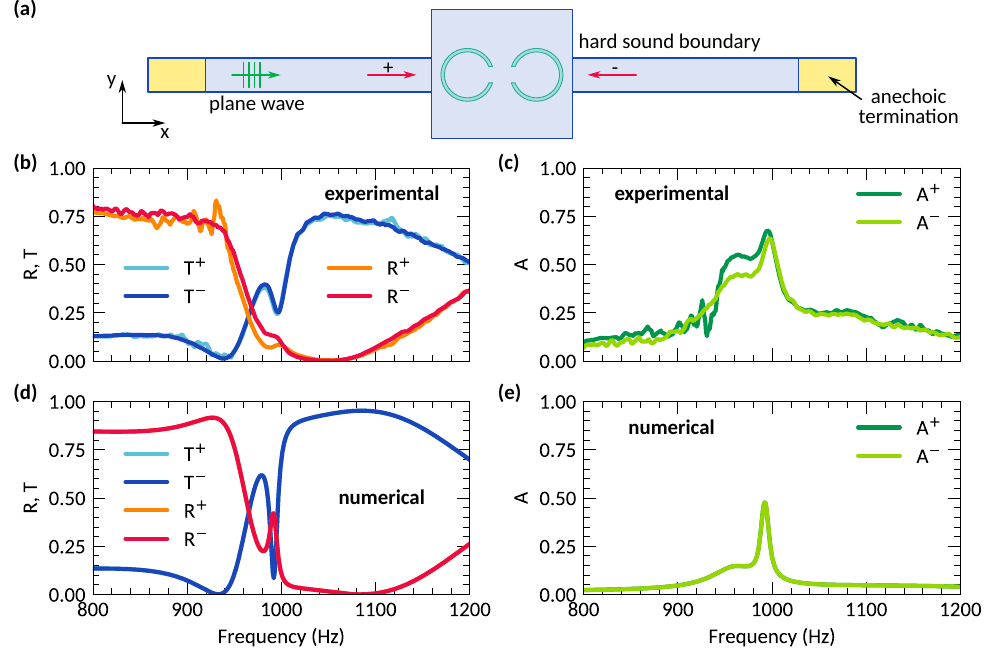}
    \caption{\textbf{Field propagation in the symmetric system.} (a) Schematic illustration of the system in which positive and negative directions of wave propagation are indicated by the corresponding arrows with the "+" and "-" labels. (b) Experimentally measured and (d) numerically calculated reflection and transmission coefficients. (c) Experimentally measured and (e) numerically calculated absorption coefficients.}
    \label{fig:tube_TL_RT}
\end{figure}

So far, the considered system was characterized by the inversion symmetry, meaning that it is isotropic with respect to the axis along which the incident field propagates. Hence, the propagation of waves is also symmetric, such that there are no differences in propagation along positive or negative directions of the $x$-axis [see Fig.~\ref{fig:tube_TL_RT}(a)]. As shown in Figs.~\ref{fig:tube_TL_RT}(b) and~\ref{fig:tube_TL_RT}(c) the corresponding "+" and "-" scattering and absorption coefficients nearly coincide. Slight differences in the vicinity of symmetric resonance can be associated with the vibrations of the box in which the resonators are placed, which is an additional mechanism of losses in the system. The corresponding numerical spectra [see Figs.~\ref{fig:tube_TL_RT}(d) and~\ref{fig:tube_TL_RT}(e)] demonstrate that "+" and "-" coefficients are indistinguishable, so the structure is perfectly symmetric, as it should be expected.

\begin{figure}[htbp!]
    \centering
    \includegraphics[width=0.9\linewidth]{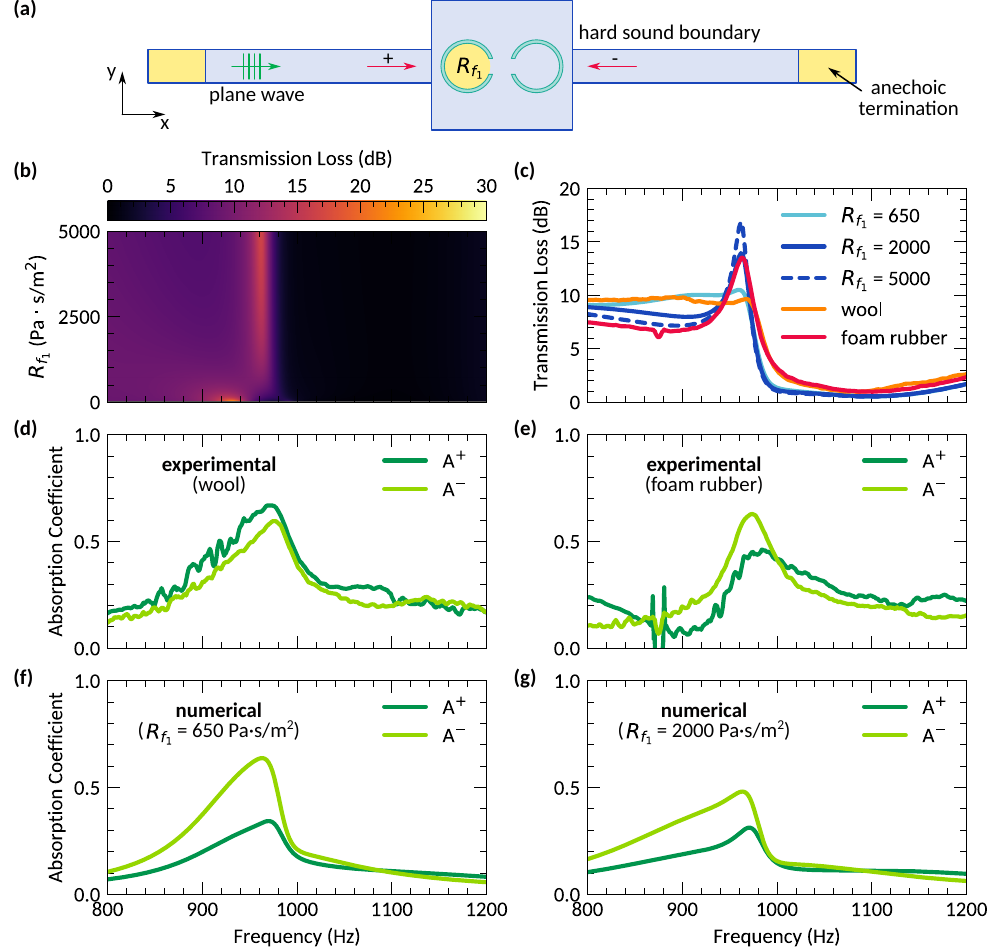}
    \caption{\textbf{Asymmetric absorption of the bi-anisotropic system.} (a) Schematic illustration of the considered system, in which one of the resonators is supplemented by a porous insert with the flow resistivity $R_{f_1}$. (b) Numerically calculated transmisison loss spectra as a function of $R_{f_1}$. (c) Comparison of transmission loss spectra measured for the inserts made of wool and foam rubber with the corresponding numerically obtained spectra. The results on panels (b) and (c) are obtained for the case when incident wave propagates along the positive direction. Experimentally measured absorption spectra for the inserts made of (d) wool and (e) foam rubber. Numerically calculated absorption spectra for the inserts with the flow rsistivity (f) $650$~Pa$\cdot$s/m$^2$ and (g) $2000$~Pa$\cdot$s/m$^2$.}
    \label{fig:tube_dx100_TL_porous_map}
\end{figure}

The inversion symmetry break can be induced via introduction of additional intrinsic losses into one of the resonators. As previously these losses can be introduced as an insert made of absorbing material, characterized by the flow resistivity $R_{f_1}$, as shown in Fig.~\ref{fig:tube_dx100_TL_porous_map}(a).
According to the numerically calculated transmission loss spectra, the increase of $R_{f_1}$ results in the destruction of one of the resonances [see Fig.~\ref{fig:tube_dx100_TL_porous_map}(b)], which can be associated with the destruction of the symmetric and antisymmetric modes, resulting in the formation of the resonance corresponding to a single resonator. Notably, at some values of $R_{f_1}$ the transmission loss is practically non-resonant, while the increase of $R_{f_1}$ results of the increase of transmission loss resonance amplitude. In order to observe same effects experimentally, two types of absorbing inserts are considered, such that one of them is made of wool and another -- from foam rubber. Material parameters of the inserts are unknown, and the estimation of their flow resistivity is done simply by finding numerically obtained spectral curves of the same forms [see Fig.~\ref{fig:tube_dx100_TL_porous_map}(c)]. Contrary to the case of the system wihout the insert, the absorption spectra in this case are different for different directions of wave propagation [see Figs.~\ref{fig:tube_dx100_TL_porous_map}(d) and~\ref{fig:tube_dx100_TL_porous_map}(e)]. While for the experimentally obtained results the difference is not large, the numerically calculated $A^+$ and $A^-$ spectra demonstrate significant asymmetry of the wave propagation, as the inversion symmetry break is associated with the bi-anisotropy of the system. Poor agreement between the numerical and experimental results can be associated with losses related to the vibrations of the box walls.

\clearpage
\section{Lumped Element Model}
Theoretical treatment of the Helmholtz resonators can be done within the acoustic impedance model in which resonators are represented as resonant circuits with lumped elements. The provided description is adaptation of  the model developed in Ref.~\cite{johansson2001theory} to the considered two-dimensional case. It should be highlighted that the adaptation is far from rigorous and rely on numerical calculations rather than on analytical derivations. Nevertheless, it allows to qualitatively describe the coupling in terms of the mutual impedance of the resonators.

\begin{figure}[htbp!]
    \centering
    \includegraphics[width=0.9\linewidth]{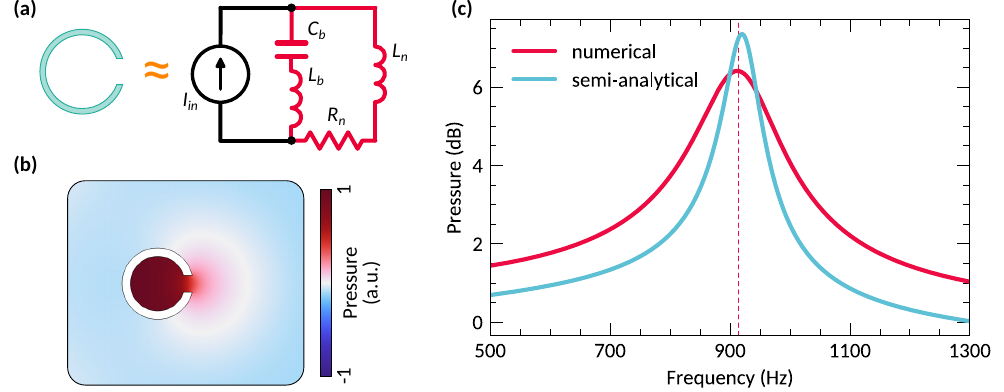}
    \caption{\textbf{Acoustic impedance model of a single Helmholtz resonator.} (a) Resonant circuit representation of the considered Helmholtz resonator. (b) Eigenmode field distribution of the Helmholtz resonator placed in a free space. (c) Frequency spectra of the pressure calculated inside the resonator excited by a plane wave and the corresponding neck volume velocity. }
    \label{fig:RLC_single}
\end{figure}

To develop the impedance model, a single Helmholtz resonator is considered to be equivalent to a simple circuit shown in Fig.~\ref{fig:RLC_single}(a), where the capacitance represents the acoustical compliance of air in the body of the resonator:
\begin{equation}
    C_b = \frac{\pi r^2}{\rho_0 c_0^2},
\end{equation}
the inductance $L_b$ is the body air mass correction,
\begin{equation}
    L_b = \frac{1}{3} \pi \rho_0,
\end{equation}
and the inductance $L_n$ corresponds to the acoustic mass of the air in the neck
\begin{equation}
    L_n = \frac{\rho_0 l_{\mathrm{eff}}}{w},
\end{equation}
such that $r$ is the inner radius of the resonator, $w$ is the neck width, and $l_{\mathrm{eff}}$ is the effective neck length, while $\rho_0$ is the density of air and $c_0$ is the speed of sound in the air. The losses are accounted for by the neck resistance $R_n$ defined as
\begin{equation}
    R_n = \frac{2l_{\mathrm{eff}}}{w} \frac{\sqrt{2 \mu_{\mathrm{eff}} \rho_0 \omega}}{w} + 2 \frac{\sqrt{2\mu_0 \rho_0 \omega}}{w} + \eta \frac{\pi \rho}{c_0} f^2
\end{equation}
where $\mu_0$ is the dynamical viscosity of air and $\mu_{\text{eff}}$ is the effective viscosity, such that
\begin{equation}
    \mu_{\text{eff}} = \mu_0 \left(1 + (1 - \gamma) \sqrt{\frac{v}{\mu_0 c_p}} \right), 
\end{equation}
where $\gamma $ is the ratio of specific heats, $v$ is thermal conductivity and $c_p$ is the heat capacity at constant pressure. Note that the first two terms of $R_n$ represent viscous losses in the neck wall boundary layer and at the neck ends. The last term corresponds to the radiation losses at the outer neck end obtained within the low-frequency limit of a piston in a long tube. It is assumed, that this term should be adjusted in order to provide a proper fitting of the model as the neck of the considered resonator can't be considered as a long tube, but still the radiation losses are of the monopole type [see Fig.~\ref{fig:RLC_single}(b)]. 
Hence, the fitting parameter $\eta$ was introduced, which was estimated to be $\SI{0.1}{\meter}$. Note, that the fitting parameter is dimensional in order to make the units of the last term of $R_n$ be consistent with the units of the two other terms.
The effective neck length was also estimated from numerical calculations by simple brute-forcing to align the resonant frequencies of pressure inside the resonator and the neck volume velocity. In particular, the estimated value is
\begin{equation}
    l_{\text{eff}} = R - r + 0.55 w.
\end{equation}
Within the considered lumped-element model, the voltage in the resonant circuit or, equivalently, the neck volume velocity of the resonator is then can be written as 
\begin{equation}
    I_n = \frac{Z_b}{Z_n + Z_b} I_{\mathrm{in}},
\end{equation}
where $Z_b$ and $Z_n$ are the body and neck impedances, such that
\begin{gather}
    Z_n = R_n + i\omega L_n,
    \\
    Z_b = \dfrac{1}{i \omega C_b}.
\end{gather}
As shown in Fig.~\ref{fig:RLC_single}(c), the spectra of the pressure inside the resonator and the corresponding neck volume velocity almost coincide, though the aim of fitting was only the alignment of resonant frequencies.

\begin{figure}[htbp!]
    \centering
    \includegraphics[width=0.9\linewidth]{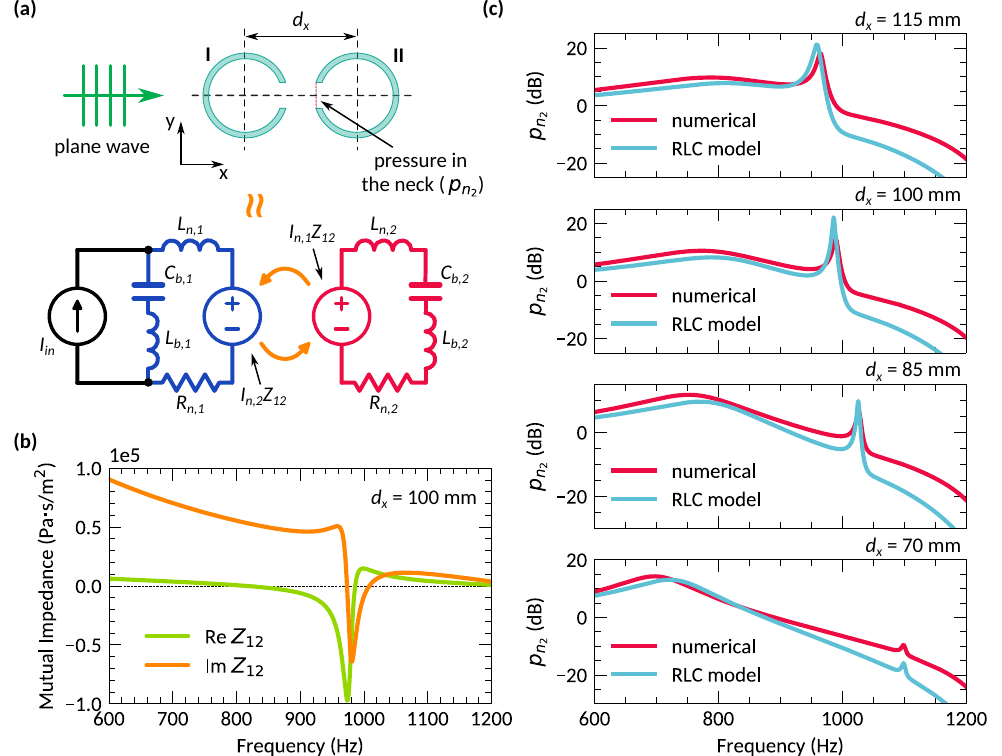}
    \caption{\textbf{Acoustic impedance model of coupled Helmholtz resonators.} (a) Resonant circuit representation of the considered coupled Helmholtz resonators.  Frequency spectra of (b) the mutual impedance and (c) the pressure calculated inside the neck of one of the resonators (labelled as II). }
    \label{fig:RLC_coupled_right}
\end{figure}

When the second resonator is introduced into the system, it can be considered as an additional resonant circuit [see Fig.~\ref{fig:RLC_coupled_right}(a)]. The radiation from one resonator results in the pressure on the neck opening of another resonator, which can be taken into account via introduction of the mutual impedance $Z_{12}$ defined as the ratio of the pressure in the neck opening of one resonator over the unit volume velocity of another resonator:
\begin{equation}
    Z_{12} = \frac{p_{12}}{I_{2}}.
\end{equation}
The mutual impedance allows to describe how one resonator acts as a source for another resonator. In addition to that, the waves are scattered from the boundaries of the resonators, which can be taken into account via introduction of the reflection impedance, defined as the pressure of the scattered field over the corresponding unit volume velocity:
\begin{equation}
    Z_{r,12} = \frac{p_{\text{sc},12}}{I_2}, ~~~ Z_{r,21} = \frac{p_{\text{sc},21}}{I_1}.
\end{equation}
Despite the fact that the mutual and reflection impedance can be estimated analytically~\cite{johansson2001theory,smith2022asymptotics}, the values of $p_{12}$ and $I_2$ are determined from the numerical calculations. 
The total neck impedance of the resonators then can be written as
\begin{align}
    Z_{n,1} = Z_n + Z_{r,12},
    \\
    Z_{n,2} = Z_n + Z_{r,21}.
\end{align}
Finally, the neck volume velocities of the resonators are
\begin{align}
    I_{n,1} = I_{\mathrm{in}} Z_{b,1} \frac{Z_2}{Z_1 Z_2 - Z_{12}^2},
    \\
    I_{n,2} = I_{\mathrm{in}} Z_{b,1} \frac{-Z_{12}}{Z_1 Z_2 - Z_{12}^2},
\end{align}
where the following notation is used:
\begin{align}
    Z_1 = Z_{n,1} + Z_{b,1},
    \\
    Z_2 = Z_{n,2} + Z_{b,2}.
\end{align}

As shown in Fig.~\ref{fig:RLC_coupled_right}(c), the neck pressure estimated using the lumped element model follow the same pattern as the numerically obtained pressure spectra, such that the spectral distance between the resonances decreases with the increase of the distance between the resonators  The pronounced resonance occurring within the $900$--$\SI{1100}{\hertz}$ range is associated with the resonance in the mutual impedance. However, it should be highlighted that the provided model can't be considered as a completely fare one as it is based on the numerical calculations and bruteforcing of some parameters. Nevertheless, it allows to provide a decent analogy of coupled resonators with the resonant circuits, which in turn highlights versatility of the fundamental features which can be observed in wave systems of different nature. 

\clearpage
\section{Gas Sensing}
\begin{figure}[htbp!]
    \centering
    \includegraphics[width=0.9\linewidth]{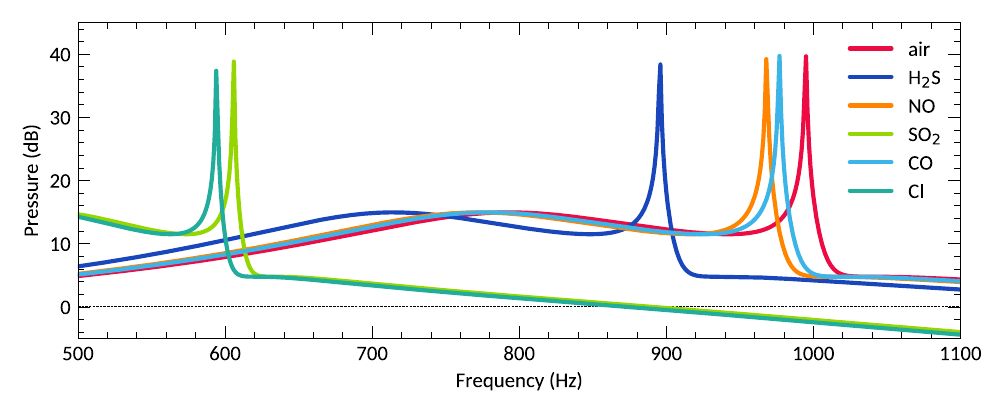}
    \caption{\textbf{Coupled resonators as gas sensors.} Pressure spectra numerically calculated inside on of the resonators placed into different gas media.}
    \label{fig:open_spectra_gases}
\end{figure}

It should be noted that the spectral position of quasi-BIC is sensitive not only to the geometric parameters of the resonators but also to the material properties of the media in which they are placed. Such property can be utilized for sensing applications in which the spectral position of a resonance is associated with the material properties of liquid of gaseous analytes~\cite{nazemi2019advanced,altug2022advances}. Typically, the higher the Q-factor and the more sensitive a resonance to the material properties of analytes, the better resolution can be achieved. The quasi-BIC resonances of the considered coupled resonator can be also utilized for the sensing purposes. As Fig.~\ref{fig:open_spectra_gases} shows, the spectral position of quasi-BIC differs for different gases in which the resonators are placed. While the development of such sensors lies far beyond the scope of the work, this result demonstrate that the developed simple system can become a basis of particular engineering applications.

\bibliographystyle{unsrt}
\bibliography{references.bib}